\documentclass[12pt,a4paper]{article}
\usepackage[tbtags]{amsmath}
\usepackage{amssymb}
\usepackage{amsthm}
\usepackage{graphics}
\usepackage{exscale}
\usepackage{ifthen}

\newcommand{\sh}[1]{
\ifthenelse{\equal{#1}{l}}
  {
   \text{$\not{#1}$}
  }
  {
   \text{$\not{\!#1}$}
  }
}

\newcommand{\intd}{\int\frac{d^dk}{(2\pi)^d}}

\title{Matching of $\lambda_B$ onto HQET}
\author{\small Volker Pilipp\footnote{volker.pilipp@physik.lmu.de}\\
\footnotesize Arnold Sommerfeld Center, Department f\"ur Physik\\
\footnotesize Ludwig-Maximilians-Universit\"at M\"unchen\\
\footnotesize Theresienstrasse 37, 80333 M\"unchen, Germany
}
\date{}

\begin{document}
\maketitle
\begin{abstract}
  The quantity $1/\lambda_B$, the inverse moment of the
  $B$-meson light-cone wave function, plays an important role in
  exclusive $B$-decays. I calculate the matching of $\lambda_B$ 
  defined in QCD onto $\lambda_B$ defined in HQET.
  This is useful for comparing results that have been obtained
  in QCD to results obtained in HQET.
\end{abstract}

\section{Introduction}
The $B$-meson wave function can be defined with two different types of
fields. Either we use the $b$-quark field, which occurs in the ordinary
QCD Lagrangian, or the heavy quark field, which occurs in the Lagrangian
of HQET \cite{georgi,HQET}. In any case we get different wave functions,
which differ at subleading order in $\alpha_s$. 
If we want to compare a QCD calculation
to the corresponding calculation, made in an effective theory like
HQET or SCET, we need the connection between the QCD and the HQET wave
functions. In this paper I will compute the matching of $\lambda_B$, the
inverse moment of the $B$-meson wave function. This parameter often
appears in exclusive $B$-decays like $B\to\pi\pi$ or $B\to \gamma l\nu$.
Usually the LO
result only depends on $\lambda_B$, while the higher logarithmic moments
appear first at NLO. It is then sufficient to know the
$\alpha_s$-corrections of the matching coefficient of
$\lambda_B$. I have tested my result by calculating the NLO of the
hard spectator scattering amplitude in $B\to\pi\pi$ in QCD \cite{pilipp}. 
The same calculation has been performed before by \cite{hsp,kivel} within 
the framework of SCET. The results agree, 
if the matching of $\lambda_B$ is properly taken into account.

\section{Definitions}
For the $B$-meson wave function I use the definition of 
\cite{grozin,laneu}:
\begin{equation}
i\hat{f}_B(\mu) m_B \phi_+^\text{HQET}(\omega,\mu)=
\frac{1}{2\pi}\int dt\,e^{i\omega t}
\langle 0|\bar{q}(z)[z,0]\sh{n}\gamma_5 h_v(0)|\bar{B}\rangle
\label{l1}
\end{equation}
and analogously for the QCD fields
\begin{equation}
if_B m_B \phi_+^\text{QCD}(\omega,\mu)=
\frac{1}{2\pi}\int dt\,e^{i\omega t}
\langle 0|\bar{q}(z)[z,0]\sh{n}\gamma_5 b(0)|\bar{B}\rangle.
\label{l2}
\end{equation}
Here $n$ is an arbitrary Lorentz vector with $n^2=0$ and $n\cdot v=1$,  
where $v$ is the four-velocity of the $B$-meson. We assume
$z\| n$. The integration in (\ref{l1}) and (\ref{l2}) 
goes over $t=v\cdot z$. The path-ordered gauge factor is given by
$[z,0]=\text{Pexp}[ig_s\int_0^1dt\,z\cdot A(tz)]$.
We define the $B$-meson decay constant by
\begin{equation}
if_Bm_B=\langle0|\bar{q}(0)\sh{n}\gamma_5b(0)|\bar{B}\rangle
\label{l3}
\end{equation}
and analogously the HQET decay constant, which depends on the
renormalisation scale $\mu$, by
\begin{equation}
i\hat{f}_B(\mu)m_B=
\langle0|\bar{q}(0)\sh{n}\gamma_5h_v(0)|\bar{B}\rangle.
\label{l4}
\end{equation}
The definition of $\lambda_B$ reads
\begin{equation}
\frac{1}{\lambda_B^\text{QCD}(\mu)}=
\int_0^\infty d\omega\,\frac{\phi_+^\text{QCD}(\omega,\mu)}{\omega}
\label{l5}
\end{equation}
and
\begin{equation}
\frac{1}{\lambda_B^\text{HQET}(\mu)}=
\int_0^\infty d\omega\,\frac{\phi_+^\text{HQET}(\omega,\mu)}{\omega}.
\label{l6}
\end{equation}
Because $\lambda_B$ usually appears in the combination
$f_B/\lambda_B$ we define our matching coefficient $C_{\lambda_B}$ 
in the following way:
\begin{equation}
\frac{f_B}{\lambda_B^\text{QCD}(\mu)}=
C_{\lambda_B}(\mu)
\frac{\hat{f}_B(\mu)}{\lambda_B^\text{HQET}(\mu)}.
\label{l7}
\end{equation}
\section{Matching calculation}
\begin{figure}
\begin{center}
\resizebox{0.7\textwidth}{!}{\includegraphics{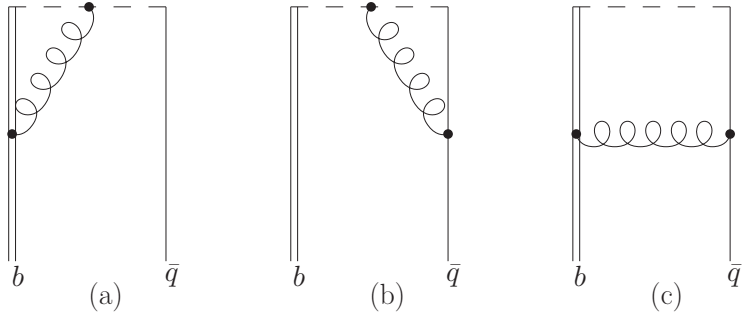}}
\end{center}
\caption{NLO contributions to $\lambda_B$. The double line stands for
 the $b$-quark field.}
\label{lbpic1}
\end{figure}
It is obvious that at LO in $\alpha_s$ and in leading power in
$\Lambda_\text{QCD}/m_b$ we get $C_{\lambda_B}=1$. 
We get the NLO correction of $C_{\lambda_B}$ by calculating the
convolution integrals over $\omega$ in (\ref{l5}) and (\ref{l6}) up to
$\mathcal{O}(\alpha_s)$. The corresponding diagrams are shown in
fig.~\ref{lbpic1}. Because $C_{\lambda_B}$ does not depend on the
hadronic physics, we use wave functions that are defined by free on-shell
quark states, i.e.\ we replace $|\bar{B}\rangle$ in (\ref{l1}) and
(\ref{l2}) by $|b(p)\bar{q}(l)\rangle$. We assign to the $b$-quark the
momentum $p=v(m_b-\tilde{\omega})$ ($-v\tilde{\omega}$ resp.) in the case of
pure QCD (HQET resp.) and $l=v\tilde{\omega}$ to the soft constituent quark,
where $v$ is the four-velocity of the $B$-meson. We assume that
$\tilde{\omega}\ll m_b$ and calculate the diagrams only in leading
power in $\tilde{\omega}/m_b$.

The diagram in fig.~\ref{lbpic1}(b) is trivially identical in QCD and
HQET, as the heavy quark field does not occur in the loop integral.
It turns out that also fig.~\ref{lbpic1}(c) does not contribute to
(\ref{l7}) in leading power. This is due to the fact that
fig.~\ref{lbpic1}(c) only contributes in leading power in the region
where the exchanged gluon is soft. In this region however QCD and 
HQET coincide. We can see this explicitly by writing down this
diagram for QCD, which reads up to constant factors:
\begin{equation}
  \intd
  \frac{\bar{q}\gamma^\nu\sh{k}\sh{n}\gamma_5(m_b\sh{v}+m_b-\sh{k})
    \gamma_\nu b}
  {k^2(k+\tilde{\omega}v)^2(k^2+2k\cdot vm_b)k\cdot n}.
\label{l7.1}  
\end{equation}
In the region, where $k$ is soft, (\ref{l7.1}) simplifies to
\begin{equation}
  \intd
  \frac{\bar{q}\sh{v}\sh{k}\sh{n}\gamma_5 b}
  {k^2(k+\tilde{\omega}v)^2 k\cdot v k\cdot n}
  \label{l7.2}
\end{equation}
where we used the equation of motion $(1-\sh{v})b=0$. Eq. (\ref{l7.2})
is actually the contribution of fig.~\ref{lbpic1}(c) in HQET.

For the diagram in fig.~\ref{lbpic1}(a)
we obtain in QCD:
\begin{equation}
\frac{\alpha_s}{4\pi}C_F
A_0
\left(
\frac{2+2\ln\frac{\tilde{\omega}}{m_b}}{\epsilon}+
4\ln\frac{\mu}{m_b}+4-\frac{\pi^2}{6}-2\ln^2\frac{\tilde{\omega}}{m_b}
+4\ln\frac{\tilde{\omega}}{m_b}\ln\frac{\mu}{m_b}
\right)
\label{l8}
\end{equation}
and in HQET
\begin{equation}
\frac{\alpha_s}{4\pi}C_F
A_0
\left(
-\frac{1}{\epsilon^2}+\frac{2\ln\frac{\tilde{\omega}}{\mu}}{\epsilon}
-2\ln^2\frac{\tilde{\omega}}{\mu}-\frac{\pi^2}{4}
\right).
\label{l9}
\end{equation}
Here 
\begin{equation}
A_0 = if_B\int_0^\infty
d\omega\,\frac{\phi_+^{(0)}(\omega)}{\omega}
\label{l9.1}
\end{equation}
denotes the LO matrix element, which is the same for
QCD and HQET. In (\ref{l8}) and (\ref{l9}) we have set the dimension
to $d=4-2\epsilon$ and redefined $\mu^2\to\mu^2\frac{e^{\gamma_E}}{4\pi}$,
which corresponds to the $\overline{\text{MS}}$-scheme.

The wave function renormalisation constants of the heavy quark fields
are given in the on-shell scheme for the QCD $b$-field:
\begin{equation}
Z_{2b}^\frac{1}{2}=1+\frac{\alpha_s}{4\pi}C_F\left(
-\frac{1}{2\epsilon}-\frac{1}{\epsilon_\text{IR}}-3\ln\frac{\mu}{m_b}-2
\right)
\label{match7}
\end{equation}
and for the HQET field $h_v$:
\begin{equation}
Z_{2h_v}^\frac{1}{2}=1+\frac{\alpha_s}{4\pi}C_F\left(
\frac{1}{\epsilon}-\frac{1}{\epsilon_\text{IR}}\right).
\label{match8}
\end{equation}
The renormalisation of the $q$-field drops out in the
matching. Diagrammatically the matching equation (\ref{l7}) reads:
\begin{equation}
Z_{2b}^\frac{1}{2}\left(
\raisebox{-.5cm}{\resizebox{1cm}{!}{\includegraphics{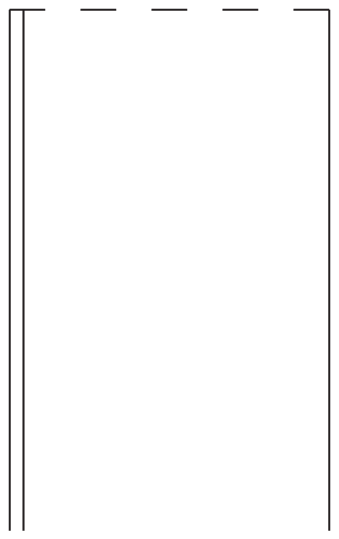}}}
+
\raisebox{-.5cm}{\resizebox{1cm}{!}{\includegraphics{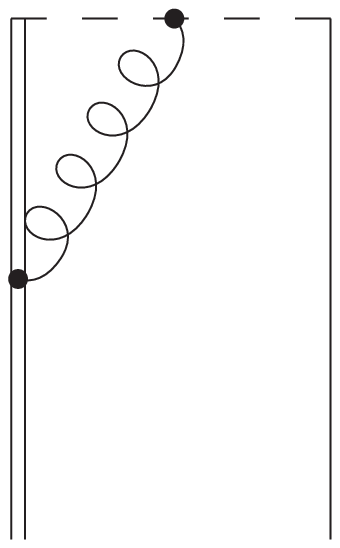}}}
\right)^\text{QCD}
=
C_{\lambda_B}
Z_{2h_v}^\frac{1}{2}\left(
\raisebox{-.5cm}{\resizebox{1cm}{!}{\includegraphics{match0}}}
+
\raisebox{-.5cm}{\resizebox{1cm}{!}{\includegraphics{match}}}
\right)^\text{HQET}.
\label{match9}
\end{equation}
Finally we obtain
\begin{equation}
C_{\lambda_B}(\mu)=1+\frac{\alpha_s}{4\pi}C_F\left(
2\ln^2\frac{\mu}{m_b}+\ln\frac{\mu}{m_b}+2+\frac{\pi^2}{12}
\right)
\label{match10}
\end{equation}
where we have renormalised the UV-divergences in the
$\overline{\text{MS}}$-scheme.

The matching coefficient for $f_B$, which has been calculated 
in \cite{HQET}, \cite{vol87}-\cite{br91}, reads: 
\begin{equation}
\hat{f}_B(\mu)=
\left(1+\frac{\alpha_s}{4\pi}C_F
\left(3\ln\frac{\mu}{m_b}+2\right)\right)
f_B.
\label{match11}
\end{equation}
This leads to:
\begin{equation}
\lambda_B^\text{HQET}=
\left(1+\frac{\alpha_s}{4\pi}C_F\left(2\ln^2\frac{\mu}{m_b}+
4\ln\frac{\mu}{m_b}+4+\frac{\pi^2}{12}\right)\right)
\lambda_B^\text{QCD}.
\label{match12}
\end{equation}
\section{Discussion}
Eq. (\ref{match12}) allows us to express the dependence of
$\lambda_B^\text{QCD}$ on
$\mu$ by the first logarithmic moment. From \cite{laneu} we get:
\begin{equation}
\frac{d}{d\ln\mu}\int_0^\infty d\omega\,
\frac{\phi_+^\text{HQET}(\omega)}{\omega}
=
C_F\frac{\alpha_s}{4\pi}\int_0^\infty d\omega\,
\frac{\phi_+^\text{HQET}(\omega)}{\omega}
\left(-4\ln\frac{\mu}{\omega}+2\right).
\label{match12.1}
\end{equation}
This leads immediately to
\begin{equation}
\frac{d}{d\ln\mu}\int_0^\infty d\omega\,
\frac{\phi_+^\text{QCD}(\omega)}{\omega}
=
C_F\frac{\alpha_s}{4\pi}\int_0^\infty d\omega\,
\frac{\phi_+^\text{QCD}(\omega)}{\omega}
\left(4\ln\frac{\omega}{m_b}+6\right).
\label{match12.2}
\end{equation}
The $\ln\mu$-term on the right hand side of (\ref{match12.1}) has
disappeared in (\ref{match12.2}). This term has been removed by the
double logarithm $\ln^2\mu$ in (\ref{match12}).
As already stated in the introduction, I calculated the hard spectator
scattering amplitude of $B\to\pi\pi$ in QCD, which has been calculated
before in \cite{hsp,kivel} in the framework of SCET. Beside the fact 
that using (\ref{match10}) makes our results coincide, it turned out that 
(\ref{match12.2}) leads to the right dependence of the amplitude on $\mu$.

There are crude approximations of $\lambda_B$ from sum rules
\cite{lambdaB,bik03,bk03}. 
In order to get an impression of the numerical implications of
(\ref{match12}) we use the value from \cite{lambdaB}:
\begin{equation}
\lambda_B^\text{QCD}(1\text{GeV})=460\pm160 \text{MeV}.
\label{match13}
\end{equation}
This leads to the numerical value of $\lambda_B^\text{HQET}$:
\begin{equation}
\lambda_B^\text{HQET}(1\text{GeV})=560\pm200 \text{MeV},
\label{match14}
\end{equation}
where $\alpha_s$ is defined by four active flavours and
$\Lambda^{\overline{\text{MS}}(4)}_\text{QCD}=325\text{MeV}$. 
We see that numerically the value of $\lambda_B^\text{HQET}$ is slightly
enhanced. However this enhancement is within the error range of
(\ref{match13}).

Instead of using sum rules $\lambda_B$ might be obtained
experimentally from radiative decays 
$B\to\gamma l\nu,\gamma ll,\gamma\gamma$. These decays have been
calculated in \cite{dgs02a,dgs02b,lpw02} at order $\alpha_s$, where
the results are given in terms of $\lambda_B^\text{HQET}$. The
corresponding $\lambda_B^\text{QCD}$ can be obtained by (\ref{match12}).

\subsubsection*{Acknowledgement}
I would like to thank Gerhard Buchalla for useful discussions and for
comments on the manuscript.

\end{document}